\documentclass[twocolumn,showpacs]{revtex4}

\usepackage{graphicx}
\usepackage{dcolumn}
\usepackage{amsmath}

\makeatletter
\def\btt#1{\texttt{\@backslashchar#1}}%
\DeclareRobustCommand\bblash{\btt{\@backslashchar}}%
\makeatother

\begin{document}

\title[]{New Superconducting and Magnetic Phases Emerge on the Verge of Antiferromagnetism in CeIn$_3$}

\author{Shinji Kawasaki,$^1$ Takeshi Mito,$^{1,*}$ Yu Kawasaki,$^{1,**}$ Hisashi Kotegawa,$^{1,***}$ Guo-qing Zheng,$^{1}$ Yoshio Kitaoka,$^{1}$ Hiroaki Shishido,$^{2}$ Shingo Araki,$^{2}$ Rikio Settai,$^{2}$ Yoshichika \=Onuki$^{2,3}$}

\affiliation{
$^1$Department of Physical Science, Graduate School of Engineering Science, Osaka University, Toyonaka, Osaka 560-8531, Japan\\$^2$Department of Physics, Graduate School of Science, Osaka University, Toyonaka, Osaka 560-0043, Japan\\$^3$Advanced Science Research Center, Japan Atomic Energy Research Institute, Tokai, Ibaraki 319-1195, Japan}

\date{\today}

\begin{abstract}

 We report the discovery of new superconducting and novel magnetic phases in CeIn$_3$ on the verge of antiferromagnetism (AFM) under pressure ($P$) through the In-nuclear quadrupole resonance (NQR) measurements. We have found a $P$-induced phase separation of AFM and paramagnetism (PM) without any trace for a quantum phase transition in CeIn$_3$. A new type of superconductivity (SC) was found in $P=2.28-2.5$ GPa to coexist with AFM that is magnetically separated from PM where the heavy fermion SC takes place. We propose that the magnetic excitations such as spin-density fluctuations induced by the first-order magnetic phase transition might mediate attractive interaction to form Cooper pairs.
\end{abstract}


                       
\maketitle
The most common kind of superconductivity (SC) is based on bound electron pairs coupled by deformation of lattice. However, SC of more subtle origins is rife in strongly correlated electrons systems including many heavy fermion (HF), cuprate and organic superconductors. Especially, a number of studies in $f$-electron compounds revealed that unconventional SC arises at or close to the quantum phase transition, i.e. the quantum critical point (QCP), where magnetic order disappears at low temperature ($T$) as a function of lattice density via application of hydrostatic pressure ($P$). These findings suggest that the mechanism forming Cooper pairs can be magnetic in origin.
Namely, on the verge of magnetic order, the magnetically soft electron liquid can mediate spin-dependent attractive interactions between the charge carriers\cite{Mathur}.  
The phase diagram, schematically shown in Fig.~1(a), has been observed in antiferromagnetic HF compounds such as CePd$_2$Si$_2$\cite{Grosche,Mathur,Grosche2}, CeIn$_3$\cite{Walker,Mathur,Grosche2,Muramatsu2,Knebel}, and CeRh$_2$Si$_2$\cite{Movshovic,Araki}. Remarkably different behavior, schematically shown in Fig.1(b), has been found in the archetypal HF superconductor CeCu$_2$Si$_2$\cite{Steglich,Bellarbi,Ykawasaki,Ykawasaki2} and the more recently discovered CeRhIn$_5$\cite{Hegger,Muramatsu}. Although an analogous behavior relevant with a magnetic QCP has been demonstrated in both the compounds, it is noteworthy that the associated superconducting region extends to higher densities than in the other compounds, their value of $T_c$ reaching its maximum away from the verge of antiferromagnetism (AFM)\cite{Bellarbi,Muramatsu}. Most interestingly, the recent nuclear quadrupole resonance (NQR) studies have unraveled that AFM and SC coexist microscopically and its SC does not yield any trace of line-node gap opening in the low-lying excitations below $T_c$ characteristic for the HF superconductors reported thus far\cite{Ykawasaki2,Mito,Skawasaki}. However, the underlying mechanism of HF SC and its relation to AFM are under debate and hence more systematic experiments are highly required.

In this paper, we report a new superconducting phase emerging on the verge of the AFM in CeIn$_3$ via the $^{115}$In-NQR measurements under $P$ and at low temperatures, the most powerful tool for the study of both AFM and SC. We have deepened the understanding of the physical properties on the verge of AFM in CeIn$_3$ that exhibits the archetypal phase diagram shown in Fig.1(a).  
We show in Fig.1(c) the $P$ - $T$ rich phase diagram in CeIn$_3$ around critical $P$($P_c$) determined from the present experiments. The phase separation of AFM and PM is evidenced in CeIn$_3$ from the observation of two kinds of NQR spectra  in $P = 2.28 - 2.50$ GPa. Nevertheless, it is highlighted that the SC in CeIn$_3$ occurs in both the phases at $P =$ 2.43 GPa, where the maximum value of $T_c^{max} = 230$ mK is observed for SC in PM and most remarkably, the SC coexisting with AFM emerges below $T_c = 190$ mK. Present results indicate the occurence of the first-order phase transition from SC coexisting with AFM to SC in HF state around $P_c$ without antiferromagnetic QCP.

CeIn$_3$ forms in the cubic AuCu$_3$ structure and orders antiferromagnetically below the N\'eel temperature $T_N = 10.2$ K at $P = 0$ with an ordering vector {\bf Q} = (1/2,1/2,1/2) and Ce magnetic moment $M_S\sim 0.5\mu_B$ which were probed by NQR measurements\cite{Kohori} and the recent neutron diffraction experiment on single crystals\cite{Knafo}, respectively. The resistivity measurements of CeIn$_3$ have clarified the $P$ - $T$ phase diagram of AFM and SC: $T_N$ decreases with increasing $P$ and on the verge of AFM, SC emerges in a narrow $P$ range of about 0.5 GPa, exhibiting a maximum value of $T_c\sim 0.2$ K around a $P_c = 2.5$ GPa where AFM disappears\cite{Walker,Mathur,Grosche2,Muramatsu2,Knebel}.
A non-Fermi-liquid behavior was suggested from the $T^{3/2}$ dependence in resistivity deviated from the conventional $T^2$ dependence in a quite narrow $P$ range around $P_c$. Then, it was inferred that the magnetically soft electron liquid can mediate spin dependent attractive interactions between the charge carriers in CeIn$_3$\cite{Walker,Mathur,Grosche2,Muramatsu2,Knebel}. 
In the previous papers\cite{Skawasaki2,Thessieu,Skawasaki3}, we reported systematic change in magnetic character via the  measurements of nuclear-spin-lattice-relaxation rate $1/T_1$ through $^{115}$In NQR under $P$: The localized magnetic character is robust up to  $P=$1.9 GPa. A characteristic temperature $T^{*}$, below which the system crosses over to an itinerant magnetic regime, increases dramatically with further increasing $P$.  As a result, the measurements of $1/T_1$ and ac-susceptibility ($\chi_{ac}$) at $P = 2.65$ GPa down to $T=50$ mK provided first evidence for unconventional SC at $T_c=95$ mK in CeIn$_3$, which occurs at the HF state fully established below
 $T_{FL} = 5$ K\cite{Skawasaki3}.


High quality single crystals of CeIn$_3$ were grown by the Czochralski method and moderately crushed into grains in order to make rf pulses penetrate into samples easily. To avoid crystal distortions, however, the size of the grains is kept with diameters larger than 100 $\mu$m. A small piece of CeIn$_3$ cut from the same batch used in the present work exhibited the zero resistance in $P = 2.2 - 2.8$ GPa\cite{Muramatsu2}, which is in good agreement with the previous reports\cite{Mathur,Grosche2,Knebel}. The NQR spectrum was obtained by plotting a spin-echo intensity as a function of frequency. The NQR $T_1$ was measured by the conventional saturation-recovery method in $T = 0.05 - 70$ K and  $P = 2.17 - 2.65$ GPa. The 2$\nu_{Q}$ ($\pm 3/2\leftrightarrow \pm 5/2$) and 1$\nu_{Q}$ ($\pm 1/2\leftrightarrow \pm 3/2$) transitions were used for the $T_1$ measurement above and below $T$ = 1.4 K, respectively.  The high frequency $\chi_{ac}$ was measured by using an {\it in-situ} NQR coil\cite{Mito}. Hydrostatic pressure was applied by utilizing a NiCrAl-BeCu piston-cylinder cell, filled with Si-based organic liquid as a pressure-transmitting medium. To calibrate the value of pressure at low temperatures, the shift in $T_c$ of Sn metal under $P$ was measured by the conventional four terminal resistivity measurement. In this experimental condition, the possible $P$ distribution is estimated to be less than $3 \%$ from a broadening in the linewidth in NQR spectrum. To reach a lowest temperature of 50 mK, a $^3$He-$^4$He dilution refrigerator was used.

Figure. 2 shows the NQR spectra of 1$\nu_{Q}$ transition for the PM at (a) : $P = 2.37$ GPa and for temperatures lower than $T_N$ and $T_c$ at (b) : $P = 2.37$ GPa, (c) : $P = 2.43$ GPa and (d) : $P = 2.50$ GPa. Note that the 1$\nu_Q$ transition can sensitively probe the appearance of internal field associated with even tiny Ce ordered moments on the verge of AFM. As a matter of fact, as seen in Figs. 2(a) and 2(b), a drastic change in the NQR spectral shape is observed due to the occurrence of internal field at the In nuclei below $T_N$. By contrast, the spectra at (c): $P = 2.43$ GPa and  (d):  $P = 2.50$ GPa include two kinds of spectra arising from AFM and PM, providing microscopic, firm evidence for the emergence of magnetic phase separation. The volume fraction of AFM at $T = $100 mK at each pressures is plotted as a function of $P$ as seen in Fig.3(a)(solid triangles). It should be noted that, as shown in Fig.1(c), the phase separations at $P$ = 2.28 and 2.37 GPa are observed only between $T$ = 3 K and $T_N$ = 5.2 K and between $T$ = 1 K and $T_N = 4.9$ K, respectively.

Figure.~3(b) shows the $T$ dependence of $\chi_{ac}$ for CeIn$_3$ under various values of $P$.  Even though the magnetic phase separation takes place as shaded in Fig.1(c), a clear decrease in $\chi_{ac}$ points to a bulk nature of superconducting transitions in $P = 2.28 - 2.65$ GPa. It is, however, noteworthy that there is no indication of SC down to $T$ = 30 mK at $P$ = 2.17 GPa where any magnetic phase separation is not observed at all against $P$ and $T$. When taking into account the fact that the value of $T_c$ reaches a maximum at $P =$ 2.43 GPa where the volume fraction of AFM and PM remains comparable, it is expected that the SC coexisting with AFM occurs only in the $P$ region where AFM and PM are separated. From the results of $P$ dependence of NQR spectrum and $\chi_{\rm ac}$, it is strongly suggested that AFM and SC coexist in the shaded region in Fig. 3(a). Also, our results suggest that a first-order transition from the coexistent phase of SC and AFM (shaded area in Fig. 3(a)) to the SC phase with paramagnetic background (area marked with SC(PM) in Fig. 3(a)).

The coexistence of AFM and SC is corroborated by a direct evidence from the $T$ dependence of $1/T_1T$ that can probe the low-lying excitations due to quasiparticles in SC and magnetic excitations in AFM.  Fig.~4 shows the drastic evolution in the $P$ and $T$ dependencies of $1/T_1T$ for AFM (solid symbols) and PM (open symbols) at $P$ = 2.17, 2.28, 2.43 and 2.50 GPa. Here $T_N$ is determined from the temperature below which the NQR intensity for PM decreases due to the emergence of AFM associated with the magnetic phase separation. The $T_1$ for AFM and PM is separately measured at respective NQR peaks which are clearly distinguished from each other as shown in Figs.~2(b)-2(d). Thus, the respective $T_c^{AFM}$ and $T_c^{PM}$ for AFM and PM are determined from the temperature below which $1/T_1T$ decreases markedly due to the SC gap opening. These results demonstrate that AFM and SC coexist in $P = 2.28 - 2.50$ GPa.

In the PM at $P$ = 2.43 and 2.50 GPa, a $1/T_1T$ = const. relation is valid below $T_{FL}\sim 3.2$ K and $\sim 3.5$ K, respectively, indicative of the Fermi-liquid state realized. This result is in good agreement with the previous resistivity measurement that confirmed the $T^2$ dependence in resistance\cite{Knebel}. Note that the $1/T_1$ for SC in PM at $P = 2.50$ GPa follows a $T^3$ dependence below $T_c = $190 mK, consistent with the line-node gap model characteristic for unconventional HF SC.

As shown in Fig.~4(a), no phase separation occurs below $T_N = 5.5$ K at $P = 2.17$ GPa and the $1/T_1T = $ const. behavior is observed well below $T_N$ as well as the result observed at $P = 0$\cite{Kohori}. The $T$ dependence of $1/T_1T$ at $P = 2.28$ GPa resembles the behavior for $P = 2.17$ GPa above $T \sim$ 1 K. However, at $P = 2.28$ GPa, the phase separation of AFM and PM occurs in the small $T$ window between $T_N = 5.2$ K and 3 K. In contrast to the $1/T_1T = $ const. behavior well below $T_N$ at $P = 2.17$ GPa, the $1/T_1T$ at $P = 2.28$ GPa continues to increase upon cooling below $T_N$ exceeding the values around $T_N$ in spite of antiferromagnetic spin polarization being induced. These results suggest that the occurrence of the magnetic phase separation is closely related to the onset of SC, yielding the large enhancement in the low-lying magnetic excitations for AFM. This feature is also seen for AFM at $P = 2.43$ GPa where $1/T_1T$ is larger than the value for PM as shown in Fig.~4(b). It is still unknown why such low-lying magnetic excitations continue to be enhanced well below $T_N$. Some spin-density fluctuations may be responsible for this feature  associated with the phase separation of AFM and PM.  In this context, CeIn$_3$ is  not in a magnetically soft electron liquid state\cite{Mathur}, but instead, the relevant magnetic excitations such as spin-density fluctuations induced by the first-order transition from AFM to PM might mediate attractive interaction.
Whatever its pairing mechanism is at $P = 2.28$ GPa where AFM is realized over the whole sample below $T =$ 3 K, the clear decrease in $1/T_1T$ and $\chi_{ac}$ provide convincing evidence for the coexistence of AFM and SC in CeIn$_3$ at $P$ = 2.28 GPa.

Further evidence for the new type of SC coexisting with AFM was from the results at $P =$ 2.43 GPa as indicated in Fig.~4(b).  At temperatures lower than the respective values of $T_c^{PM}$ = 230 mK and $T_c^{AFM}$ = 190 mK for PM and AFM, unexpectedly, the magnitude of $1/T_1T = $ const. coincides with one another, nevertheless both the phases are magnetically separated and the value of $T_c$ differs. This means that the low-lying excitations may be the same in origin for the SC coexisting with AFM and SC in PM. How does this case happen? It may be possible that both the phases are in a dynamically separated regime with time scales smaller than the inverse of NQR frequency so as to make each superconducting phase for AFM and PM uniform. In this context, the observed magnetically separated  phases and the relevant SC coexisting with AFM may belong to new phases of matter.

In conclusion, through the extensive $^{115}$In-NQR measurements at low temperatures under hydrostatic pressure, we have provided evidences for the phase separation of AFM and PM (shaded area in Fig. 1(c)) and new type of SC coexisting with AFM near $P_c$ in CeIn$_3$. It has been found that the highest value of $T_c = 230$ mK in CeIn$_3$ is observed for PM at $P = 2.43$ GPa where the volume fraction of AFM and PM becomes almost the same. The present experiments have revealed that the new type of SC coexisting with AFM is mediated by a novel pairing interaction relevant with the magnetic phase separation. We propose that the magnetic excitations such as spin-density fluctuations induced by the first-order magnetic phase transition might mediate attractive interaction to form Cooper pairs in CeIn$_3$- indeed a new type of pairing mechanism.
We believe that the present result will be a clue to understand the mechanism of SC in strongly correlated electron systems.

We wish to thank K. Ishida, C. Thessieu, A. V. Balatsky, T. C. Kobayashi, J. Flouquet and M. Imada for their useful discussions. One of the authors (S. K.) also thanks A. V. Kornilov for guiding his high-pressure technique.
This work was supported by Grant-in-Aid for Creative Scientific Researchi15GS0123), MEXT and The 21st Century COE Program supported by Japan Society for 
the Promotion of Science. S. K. has been supported by Research Fellowship of the Japan Society for the Promotion of Science for Young Scientists.



\noindent
* Present address: Department of Physics, Faculty of Science, Kobe University, Nada, Kobe 657-8501, Japan\\
\noindent
** Present address: Department of Physics, Faculty of Engineering, Tokushima University, Tokushima 770-8506, Japan\\
\noindent
*** Present address: Department of Physics, Faculty of Science, Okayama University, Okayama 700-8530, Japan\\

\clearpage
\begin{figure}[h]
\caption[]{Schematic phase diagrams of HF compounds: (a) for CePd$_2$Si$_2$\cite{Grosche,Mathur,Grosche2}, CeIn$_3$\cite{Walker,Mathur,Grosche2,Muramatsu2,Knebel} and CeRh$_2$Si$_2$\cite{Movshovic,Araki}: (b) for CeCu$_2$Si$_2$\cite{Steglich,Bellarbi,Ykawasaki,Ykawasaki2} and CeRhIn$_5$\cite{Hegger,Muramatsu}. Dotted and solid lines indicate $P$ dependence of $T_N$ and $T_c$, respectively. (c)The $P$ - $T$ phase diagram of CeIn$_3$ determined from the present experiment. The $P$ and $T$ ranges where the phase separation of AFM and PM occurs is shaded in the figure.}
\end{figure}

\begin{figure}[h]
\caption[]{The $P$ dependence of $^{115}$In NQR spectrum for CeIn$_3$ at $P=2.37$ GPa above $T_N$ (a) and $P$ = 2.37 GPa (b), 2.43 GPa (c) and 2.50 GPa (d) at temperatures lower than the $T_N$ and $T_c$. The dotted line indicates a peak position at which the NQR spectrum is observed for PM.}
\end{figure}

\begin{figure}[htbp]
\caption[]{(a) The $P$ dependence of the volume fraction of AFM(solid triangle) and SC in PM(open circle). The volume fraction of AFM is evaluated from the integration of the NQR spectrum over the frequency at $T$ = 100 mK at each pressures. The superconducting volume fraction at each pressure is estimated from comparing the value of $\chi_{ac}$ at a lowest temperature for CeIn$_3$ with that for the HF superconductor CeIrIn$_5$\cite{Skawasaki3}. The phase separation of AFM and PM takes place in $P=2.28-2.50$ GPa. Note that SC coexists with AFM in shaded region. (b) The $T$ dependence of $\chi_{ac}$ in $P = 2.17 - 2.65$ GPa. The solid arrow indicates the onset $T_c$ of SC.
}
\end{figure}

\begin{figure}[htbp]
\caption[]{The $T$ dependence of $ ^{115}(1/T_1T)$ in CeIn$_3$ at $P = 2.17$ and 2.28 GPa (a), 2.43 GPa (b) and 2.50 GPa (c). Open and solid symbols indicate the data for PM and AFM measured at $\sim$ 9.8 MHz and $\sim$ 8.2 MHz, respectively. The solid arrow indicates the respective superconducting transition temperature $T_c^{PM}$ and $T_c^{AFM}$ for PM and AFM.  The dotted and dashed arrows indicate, respectively, the $T_N$ and the characteristic temperature $T_{FL}$ below which the $T_1T = $const. law (dotted line) is valid, characteristic for the Fermi-liquid state.}
\end{figure}

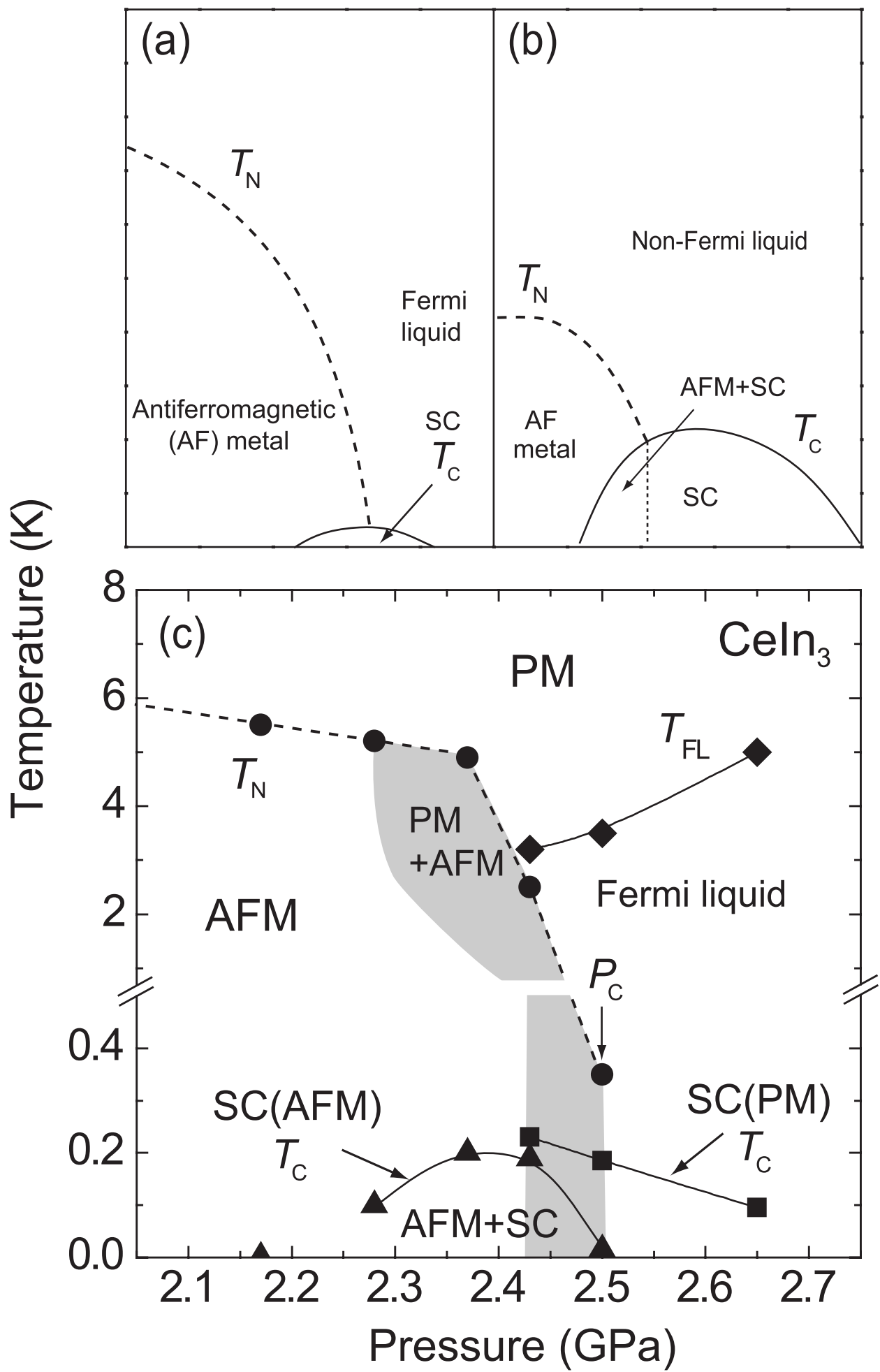
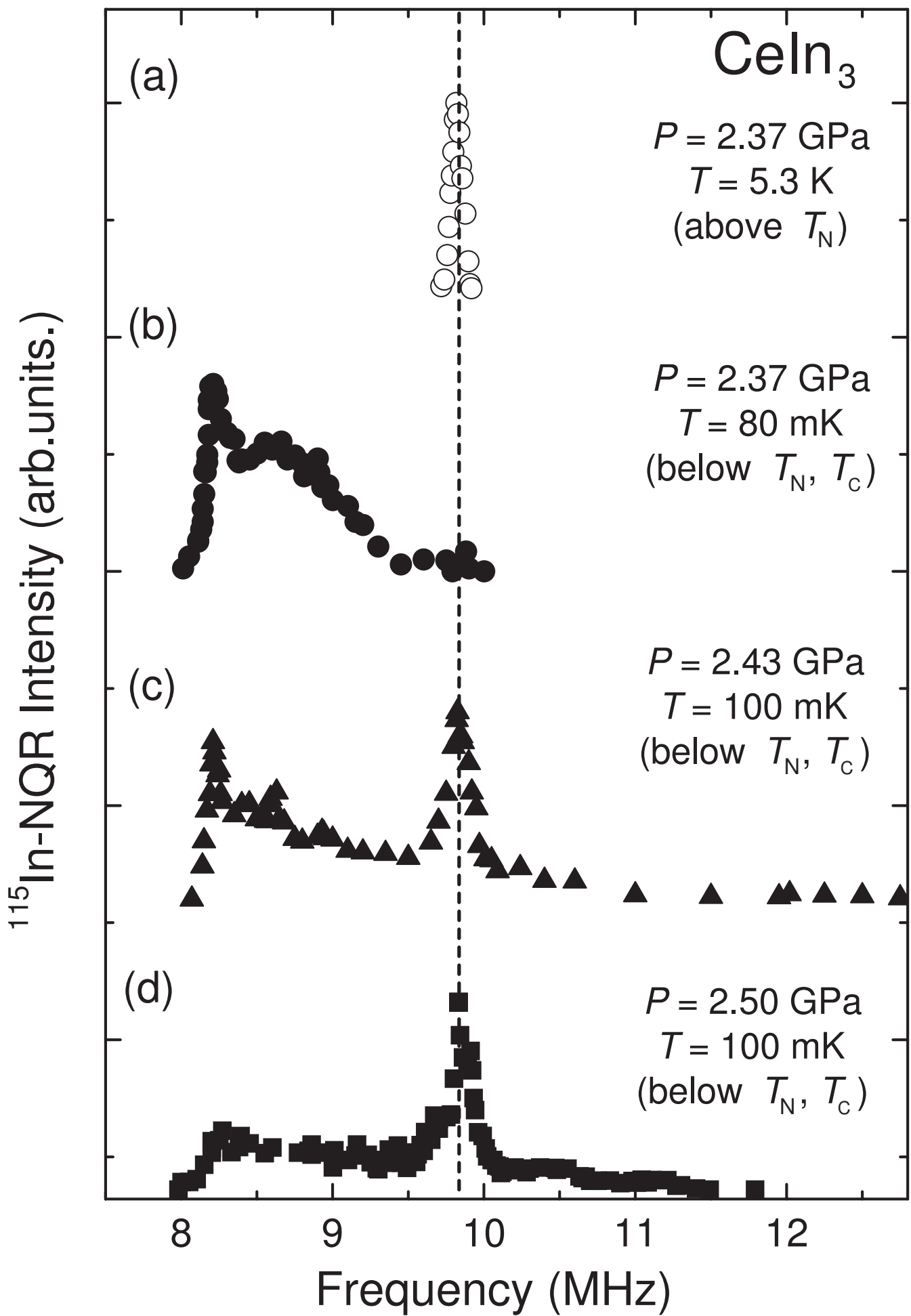
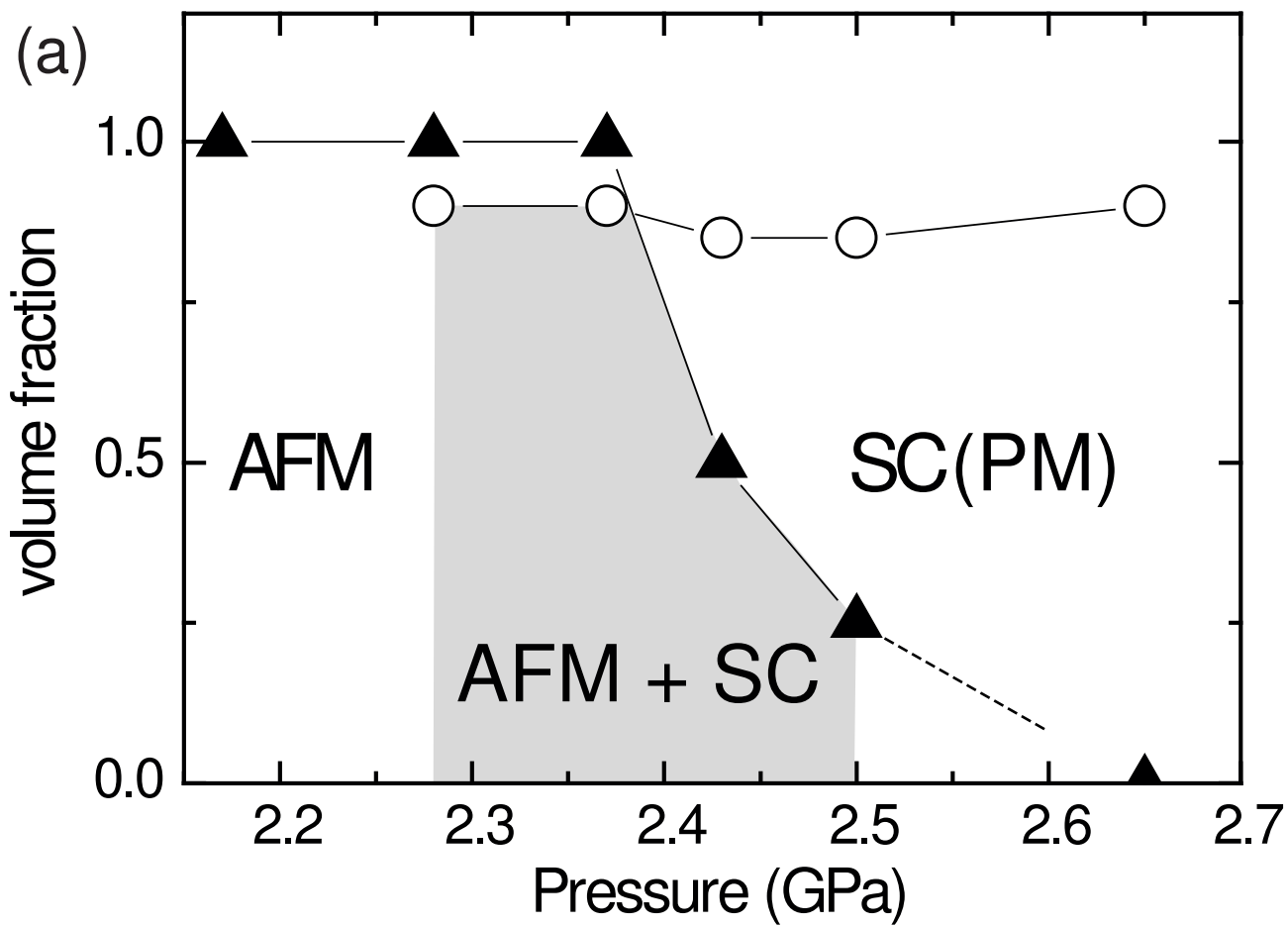
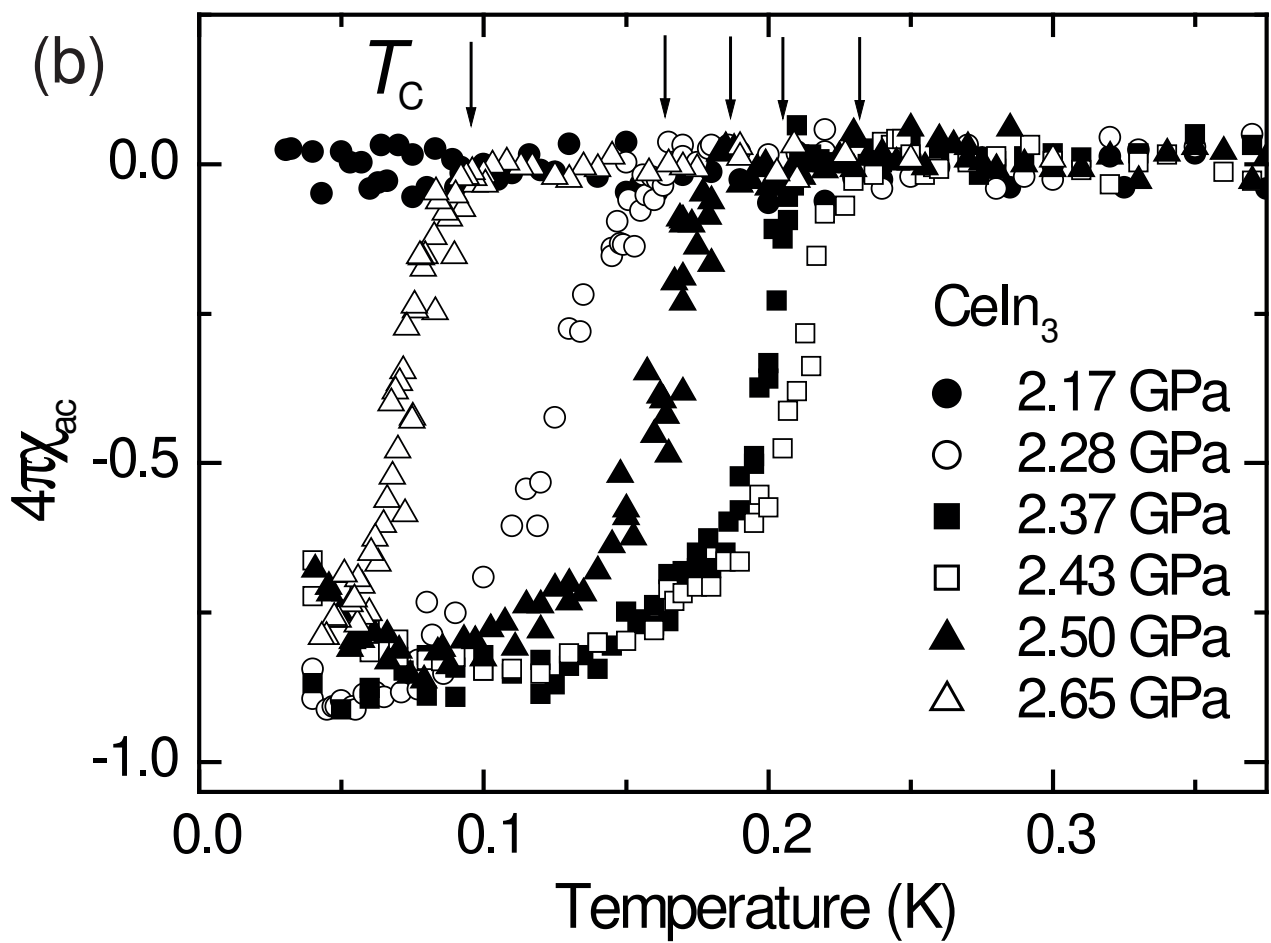
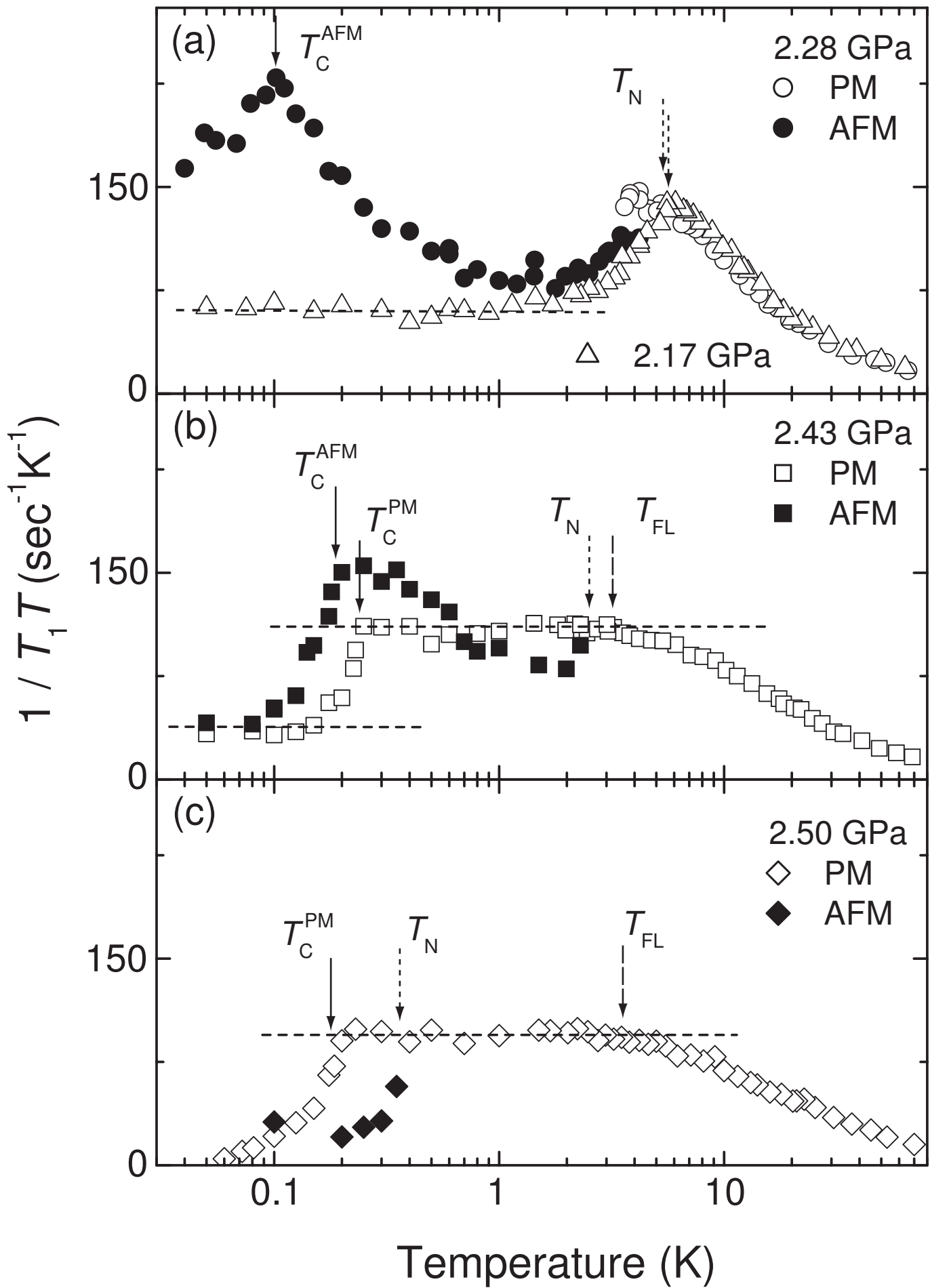
\end{document}